# Dynamic behaviors of the hexagonal Ising nanowire

**Mehmet Ertaş[1, *] and Yusuf Kocakaplan[2]**

[1] *Department of Physics, Erciyes University, 38039 Kayseri, Turkey*
[2] *Graduate School of Natural and Applied Sciences, Erciyes University, 38039 Kayseri, Turkey*

---

**Abstract**

By utilizing the effective-field theory based on the Glauber-type stochastic dynamics, the dynamic behaviors of the hexagonal Ising nanowire (HIN) system in the presence of a time dependent magnetic field are obtained. The time variations of average order parameters and the thermal behavior of the dynamic order parameters are studied to analyze the nature of transitions and to obtain the dynamic phase transition points. The dynamic phase diagrams are introduced in the plane of the reduced temperature versus magnetic field amplitude. The dynamic phase diagrams exhibit coexistence phase region, several ordered phases and critical point as well as a reentrant behavior.

*Keywords:* Magnetic nanostructured materials; Hexagonal Ising nanowire; Dynamic phase diagram; Effective-field theory; Glauber-type stochastic dynamics

---

## 1. Introduction

Recent years, magnetic nanoparticle systems such as nanowire, nanotube, nanofilms and nanorods have been receiving considerable attention by the experimental and theoretical researchers [1-7]. Because, compared with those in bulk materials, these magnetic nanomaterials have many peculiar physical properties [8]. Another reason is that they have important potential technological applications in various areas, e.g. they can be used for medical applications [9, 10], environmental remediation [11], permanent magnets [12], sensors [13], magnetic recording media [14–16], nonlinear optics [17], biotechnology [18, 19] and also used for bio-separation [20, 21].

[*] Corresponding author.
Tel: + 90 352 2076666 # 33134
*E-mail address:* mehmetertas@erciyes.edu.tr (Mehmet Ertaş)

On the other hand, many researchers have used the spin-1/2 Ising system to study equilibrium properties of magnetic nanostructured materials. An early attempt to examine nanoparticles was done by Kaneyoshi [22]. Kaneyoshi investigated the phase diagrams of a ferroelectric nanoparticle described with the transverse Ising model by using the mean-field theory (MFT) and the effective-field theory (EFT) corresponding to the Zernike approximation. Kaneyoshi also [23] studied, via the MFT, the thermal variations of longitudinal and transverse magnetizations for a ferroelectric nanoparticle system by the transverse Ising model. MFT and EFT were used to study the magnetizations [24] and phase diagrams [25] of a transverse Ising nanowire and found that the equilibrium behavior of the system is strongly affected by the surface situations. Kaneyoshi [26] also examined magnetic properties, such as the thermal behavior of total magnetization, compensation temperature, phase diagrams, total susceptibility and inverse total susceptibility, of a cylindrical spin-1/2 Ising nanowire (or nanotube) in detail within the EFT. He investigated phase diagrams and magnetizations in the cylindrical nanowire system with a diluted surface described by the transverse spin-1/2 Ising model [27] via the EFT. Keskin *et al.* [28] studied the hysteresis behaviors of the cylindrical Ising nanowire at temperatures below, around, and above the critical temperature within EFT with correlations. For ferromagnetic and antiferromagnetic interactions between the shell and the core, they obtained the hysteresis curves for different reduced temperatures. The effects of the randomly distributed magnetic field on the phase diagrams of a spin-1/2 Ising nanowire have studied by Akıncı [29] via the EFT. For the ferroelectric and anti-ferroelectric interfacial coupling, the hysteresis behaviors of the nanotube in which consisting of a ferroelectric core of spin-1/2 surrounded by a ferroelectric shell of spin-1/2 have examined by Zaim *et al.* with the transverse Ising model [30]. They obtained a number of characteristic behaviors such as the existence of triple hysteresis loops for appropriate values of the system parameters. By using the EFT based on the probability distribution technique, Bouhou *et al.* [31] investigated the magnetization, susceptibility, and hysteresis loops of a magnetic nanowire. In particular, the effects of the exchange interaction between core/shell and the external fields on the magnetization and the susceptibility were examined. They also described the diluted magnetic nanowire by the spin-1/2 Ising model and examined magnetic properties [32]. By using the EFT, Yüksel *et al.* [33] examined the effects of bond dilution on the magnetic properties of a cylindrical Ising nanowire. Very recently, Kaneyoshi studied nanoscaled thin films by the spin-1/2 Ising model and presented phase diagrams [34] and the thermal behavior of total magnetization and reentrant phenomena in a transverse Ising nanowire (or nanotube) with a diluted surface [35].

We should also mention that although the spin-1/2 Ising systems have used to investigate the equilibrium properties of magnetic nanostructured materials, there have been only a few works that the spin-1/2 Ising systems used to investigate dynamic magnetic properties of nanostructured materials. Wang *et al.* [36] studied the dynamic properties of the phase diagram in cylindrical ferroelectric nanotubes, especially they showed effects structure factors of the ferroelectric nanotubes in phase diagrams. Deviren and Keskin [37] investigated thermal behavior of dynamic magnetizations, hysteresis loop areas and cerrelations of a cylindrical Ising nanotube in a time dependent magnetic field within the EFT based on Glauber-type stochastic dynamics [38], namely dynamic effective-field theory (DEFT). Yüksel *et al.* [39] analyzed the nonequilibrium phase transition properties and hysteretic behavior of a ferromagnetic core-shell nanoparticle in the presence of a oscillating magnetic field by means of the MC simulations. Deviren *et al.* [40-42] investigated dynamic phase transitions temperature of cylindrical Ising nanowire [40], transverse cylindrical Ising nanowire [41] and the dynamic phase diagrams of the kinetic cylindrical Ising nanotube [42] using DEFT, in detail.

As far as we know the dynamics behaviors of the HIN system have not been studied yet. To overcome this deficiency in the literature, therefore, the aim of this paper is to investigate the dynamic phase transition temperatures and dynamic phase diagrams of the spin-1/2 HIN system in an oscillating magnetic field within the DEFT. We employ the Glauber-type stochastic dynamics to construct the dynamic effective-field equation for the average magnetizations. We investigate the stationary solutions of the set of coupled dynamic effective-field equations and this investigation leads us to find the phases in the system. In order to characterize the nature (continuous and discontinuous) of the phase transitions and obtain the DPT points, the thermal behavior of the dynamic order parameters have also studied. Finally, the dynamic phase diagrams in the plane of the reduced temperature versus magnetic field amplitude have presented.

The organization of the remaining part of this paper is as follows. In Section 2, the model are defined and given briefly the formulation of the spin-1/2 HIN system. In Section 3, we present the numerical results and discussions. Finally, Section 4 contains the summary and conclusions.

## 2. Model and formulation

The schematic presentation of the HIN system with core/shell structure is illustrated in Fig. 1; hence the blue and red spheres indicate magnetic atoms at the surface shell and core, respectively. Each site of the HIN system is occupied by a spin-1/2 magnetic atoms and each magnetic atom is connected to the two nearest-neighbor magnetic atoms on the above and below sections along the wire. The Hamiltonian of the system is given by

$$H = -J_S \sum_{\langle ij \rangle} S_i S_j - J_C \sum_{\langle mn \rangle} S_m S_n - J_1 \sum_{\langle im \rangle} S_i S_m - h(t)\left( \sum_i S_i + \sum_m S_m \right), \qquad (1)$$

where $<ij>$, $<mn>$ and $<im>$ stand for the summations over all pairs of neighboring spins at the shell surface, core and between shell surface and core, respectively. $S_i$ is the Pauli spin operator and it takes value $\mp 1/2$. $h(t)$ is a time-dependent external oscillating magnetic field that is given by $h(t) = h_0 \sin(\omega t)$ and $h_0$ and $\omega = 2\pi\nu$ are the amplitude and the angular frequency of the oscillating field, respectively. The $J_S$ and $J_C$ are the interaction parameters between two nearest-neighbor magnetic atoms at the surface shell and core, respectively, and $J_1$ is the interaction parameters between two nearest-neighbor magnetic atoms at the surface shell and the core. In order to clear up of the surface effects on the physical properties of the system, $J_S$ surface interaction parameter is often defined as $J_S = J_C(1+\Delta_S)$ [26, 28, 41, 42].

In the framework of the EFT with correlations by using the van der Waerden identity [43] we can easily obtain $m_C$ magnetization at the core and $m_S$ magnetization in the shell surface for the HIN system depicted in Fig. 1 as follow:

$$m_C = \left\langle \left( \prod_{i=1}^{2} \left[ \cosh(J_C \nabla) + S_i \sinh(J_C \nabla) \right] \right) \left( \prod_{j=1}^{6} \left[ \cosh(J_1 \nabla) + S_j \sinh(J_1 \nabla) \right] \right) \right\rangle F_C(x+h)\Big|_{x=0}, \qquad (2a)$$

$$m_S = \left\langle \left( \prod_{i=1}^{4} \left[ \cosh(J_S\nabla) + S_i \sinh(J_S\nabla) \right] \right) \left( \prod_{j=1}^{1} \left[ \cosh(J_1\nabla) + S_j \sinh(J_1\nabla) \right] \right) \right\rangle F_S\left(x+h\right)\Big|_{x=0}, \qquad (2b)$$

where $\nabla = \partial / \partial x$ is the differential operator. The $F_C(x+h)$ and $F_S(x+h)$ functions are defined as follow;

$$F_C(x+h) = F_S(x+h) = \frac{1}{2}\tanh\left[\frac{1}{2}\beta\left(x+h\right)\right]. \qquad (3)$$

In Eq. (3), $\beta = 1/k_B T$ where $k_B$ is the Boltzmann constant and T is absolute temperature. Moreover, the total magnetization of per site in the HIN system can be obtain via $m_T = 1/7\left(m_C + 6m_S\right)$. Also, for the following discussions, at this point let us define the r parameter as $r = J_1 / J_C$.

In order to obtain the dynamic effective-field theory (DEFT) equations of motion for the $m_C$ magnetization at the core and $m_S$ magnetization in the shell surface, we apply the Glauber-type stochastic process [38] based on the master equation as follows;

$$\frac{dm_C}{dt} = -m_C + \left[\cosh(J_C\nabla) + m_C \sinh(J_C\nabla)\right]^2 \left[\cosh(J_1\nabla) + m_S \sinh(J_1\nabla)\right]^6 F_C\left(x+h\right)\Big|_{x=0},$$

(4a)

$$\frac{dm_S}{dt} = -m_S + \left[\cosh(J_S\nabla) + m_S \sinh(J_S\nabla)\right]^4 \left[\cosh(J_1\nabla) + m_C \sinh(J_1\nabla)\right] F_S\left(x+h\right)\Big|_{x=0}.$$

(4b)

By expanding the right-hand sides of Eq. (4), and by applying the $\exp(\alpha\nabla)F(x) = F(x+\alpha)$ expression, we can obtain many coefficients. Due to their being so long and complicated, these coefficients will not be given here. The solution of Eq. (4) gives the phases in the system, and these will be presented and discussed in the next section

## 3. Numerical results and discussions

### *3.1. Phases in this system*

In this subsection, at first the time variations of the average shell and core magnetizations are examined to obtain the phases in the system. In order to investigated the behaviors of time variations of the average magnetizations, the stationary solutions of the set of coupled Eqs. (4a)-(4b) dynamic effective-field equations have been studied for different interaction parameter values. The stationary solutions of solutions of these equations will be a periodic function of $\xi$ with period $2\pi$; that is, $m_S(\xi+2\pi)=m_S(\xi)$ and $m_C(\xi+2\pi)=m_C(\xi)$. Furthermore, they can be one of the three types according to whether they have or do not have the properties

$$m_S(\xi+\pi)=-m_S(\xi), \tag{5a}$$

and

$$m_C(\xi+\pi)=-m_C(\xi). \tag{5b}$$

where $\xi=\omega t$. The first type of solution satisfies Eqs. (5a) and (5b) which is known as a symmetric solution which corresponds to a paramagnetic phase. In the solution, average magnetizations delayed with respect to the external magnetic field. The second type of solution does not satisfy Eqs. (5a) and (5b), and is called a non-symmetric solution that corresponds to a ferromagnetic (f) or antiferromagnetic (af) solution. In this case, average magnetizations do not follow the external magnetic field any more, but instead of oscillating around zero value. The third type of solution, which satisfies Eq. (5a) but does not satisfy Eq. (5b), corresponds to a nonmagnetic solution (nm). These facts are seen obviously by solving Eqs. (4a) and (4b) utilizing the Adams-Moulton predictor-corrector method for a given set of parameters and initial values, and obtained fundamental phases are presented in Fig. 2. Fig. 2 (a)-(d) illustrate paramagnetic, ferromagnetic, antiferromagnetic and nonmagnetic fundamental phases for different interaction parameters and initial values, respectivelly. In these figures each curve shows different one initial values, namely $m_C = 0.5$ and $m_S = 0.5$ or $m_C = 0.5$ and $m_S = -0.5$, and $m_C = -0.5$ and $m_S = 0.5$ etc. In Fig. 2(a), average magnetizations are equal to each other and oscillate around zero value ( $m_C(\xi)=m_S(\xi)=0$ ). Hence, the system shows symetric solution, namely paramagnetic phase. In Fig. 2(b), average shell and core magnetizations are equal to each other and oscillate around $m_C(\xi)=m_S(\xi)=0.5$ and the system illustrates ferromagnetic phase. In Fig. 2(c), as shell magnetization oscillates around 0.5 value, core magnetization oscillates around -0.5 value and system illustrates antiferromagnetic phase. In Fig. 2(d), core magnetization oscillates around the zero value and is delayed with respect to the external magnetic field and shell magnetization does not follow the external magnetic field anymore, but instead of oscillating around a zero value, it oscillates around a nonzero value, $m_S(\xi)=0.5.$ These four solutions do not depend on the initial values, seen in Fig. 2(a)-(d) explicitly. In Fig. 2(e), we have two solutions, namely nm and p phases or solutions coexist in the system and in this case, the solution depend on the initial values.

*3.2. Dynamic phase transition points*

The dynamic shell and core magnetizations or dynamic order parameters as the time-averaged magnetization over a period of the oscillating magnetic field are given as

$$M_S=\frac{1}{2\pi}\int_0^{2\pi} m_S(\xi)\,d\xi \tag{6a}$$

$$M_C=\frac{1}{2\pi}\int_0^{2\pi} m_C(\xi)\,d\xi\ . \tag{6b}$$

By combination of the Adams-Moulton predictor corrector with Romberg integration numerical methods, we solve Eqs. (6a) and (6b) and examine the thermal behavior of $M_S$ and $M_C$ for different values of Hamiltonian parameters. The thermal behaviors of $M_S$ and $M_C$ gives the dynamic phase transition (DPT) point and the type of the dynamic phase transition. The obtained numerical results of Eqs. (6a) and (6b) are seen in Figs. 3(a)-(d). In Fig. 3, $T_C$ and $T_t$ show the critical or the second-order phase transitions and the first-order phase transition temperatures, respectively. Fig. 3(a) shows the behavior of dynamic shell $M_S$ and dynamic core magnetization $M_C$ as a function of temperature for $\omega=2\pi,\ r=1.0, \Delta_S=0.0$ and $h_0$ = 0.1 values. At zero temperature, $M_S$ = $M_C$ = 0.5 and with the increase of temperature they decrease to zero continuously; thus the system undergoes a second order phase transition from the ferromagnetic phase to the paramagnetic phase at $T_C$ = 1.25. For $\omega=2\pi,\ r=-0.5, \Delta_S=-0.6$ and $h_0$ = 0.12 values, the dynamic behavior of $M_S$ and $M_C$ was obtained in Fig. 3(b). In this figure, $M_S$ and $M_C$ take 0.5 and -0.5 values at zero temperature, and they exhibits a discontinuous jump to zero from these values. Hence, the system undergoes a first-order phase transition from the antiferromagnetic phase to the paramagnetic phase at $T_t = 0.6$. Moreover, Figs. 3(c) and (d) are plotted for $\omega=2\pi$, $r=0.1$, $\Delta_S=3.0$, $h_0 = 6.05$ and different initial values, namely Fig. 3(c) for $M_C = 0.5$, $M_S = 0.5$ or $M_C = 0.5$, $M_S = -0.5$ and Fig 3(d) for $M_C = 0.0$, $M_S = 0.0$ values. In Fig. 3(c), as the temperature values increase, $M_C$ always becomes zero. In contrast, $M_S$ take the 0.5 value at zero temperature and with the increase of temperature, it undergoes a first-order phase transition at $T_t = 0.1$. Above $T_t$, $M_S$ and $M_C$ always equal to zero. On the other hand, the initial values of $M_S$ and $M_C$ are taken 0.0, they are equal to 0.0 at zero temperature and with the increase of temperature values, and they always become zero as seen in Fig. 3(d). Therefore, the system undergoes a first-order phase transition from the nm + p mixed phase to the paramagnetic phase at $T_t = 0.1$. This fact is seen in the phase diagram of Fig. 4(e) for $h_0 = 6.05$, explicitly. On the other hand, in the past two decades, both experimental (see [44-48] and references therein) and theoretical (see [49-52] and references therein) investigations of the nonequilibrium critical phenomena, especially the DPT, have received a great deal of attention due to the reason that besides the scientific interests the study of DPT can also inspired new methods in materials and manufacturing process and processing as well as in nanotechnology [53].

### *3.3. Dynamic phase diagrams*

Now, we can obtain the dynamic phase diagrams of the system. The dynamic phase diagrams are represented in the (h, T) planes for diverse values of the Hamiltonian parameters as seen in Fig. 4. In Fig. 4, the solid and dashed lines stand for the second- and first-order phase transition lines, respectively; the dynamic tricritical point (TCP) is represented by a filled circle. The dynamic phase diagrams contain the p, f, af, nm fundamental phases and the nm + p mixed phase. The system illustrates the TCP in Figs. 4(c)-4(h), but the system does not exhibit TCP in Figs. 4(a) and 4(b). In Fig. 4(a), we can see that the system always undergoes a second-order phase transition and contain dynamic zero temperature (Z) point for the $r=0.3$ and $\Delta_S=-0.95$ values. Fig. 4(b) is obtained for $r=4.0$ and $\Delta_S=-0.5$ values. It is similar to Fig. 4(a), except that the reentrant behavior is also observed in Fig. 4(b). With the temperature increase, the system passes from the p phase to the f phase, and then back to the p phase again. We should also mention that several weakly frustrated ferromagnets, such as in manganite $LaSr_2Mn_2O_7$ by electron and x-ray diffraction, in the bulk bicrystals of the oxide superconductor $BaPb_{1-x}Bi_xO_3$ and $Eu_xSr_{1-x}S$ and amorphous-$Fe_{1-x}Mn_x$, demonstrate the reentrant phenomena [54]. Fig. 4(c) and 4(d) are calculated for $r=1.0$ and $\Delta_S=0.0$, and $r=-1.0$ and $\Delta_S=0.0$, respectively. The overall structure of Fig. 4(c) and (d) is similar. Both of them contain one TCP. But, Fig. 4(c) include the f and p phases while Fig. 4(d) comprise the af and p phases. Fig. 4(e) is plotted for $r=0.1$ and $\Delta_S=3.0$, and different initial values. We can clearly see that it contains the second- and first-order phase transitions, the TCP, the Z special point and the f, p, nm and nm + p phases. For $r=0.2$ and $\Delta_S=-5.0$, besides the Z special point, nm and p phase and the TCP, Fig. 4(f) is also include reentrant behavior. Fig. 4(g) is similar to Fig. 4(f), except that the critical temperature is small. Finally, Fig. 4(h) is obtained for $r=-0.5$ and $\Delta_S=-0.6$ and it contain two TCP, and the af and p phases.

## 4. Summary and conclusions

The dynamic phase transition points (DPTs) and dynamic phase diagrams of the HIN system under a time oscillating longitudinal magnetic field were investigated using the EFT with correlations. By utilizing the Glauber-type stochastic process, the EFT equations of motion for the average shell and core magnetizations are obtained for the HIN system. Our results show that the dynamic phase diagrams contain the p, f, af and nm fundamental phases and the nm + p mixed phase as well as special Z and TCP points. We should also mention that both experimental and theoretical investigations of the nonequilibrium critical phenomena, especially the DPT, have received a great deal of attention due to the reason that besides the scientific interests the study of DPT can also inspired new methods in materials and manufacturing process and processing as well as in nanotechnology. Moreover, our results also show the reentrant behaviors and several weakly frustrated ferromagnets, such as in manganite $LaSr_2Mn_2O_7$ by electron and x-ray diffraction, in the bulk bicrystals of the oxide superconductor $BaPb_{1-x}Bi_xO_3$ and $Eu_xSr_{1-x}S$ and amorphous-$Fe_{1-x}Mn_x$, demonstrate the reentrant phenomena. Finaly, although the equilibrium phase transitions of the nanosystems have been conspicuously studied, the dynamic or nonequilibrium properties of these systems have not been investigated considerably. Hence, we hope that our work contributes to close this shortcoming in the literature.

## List of the Figure Captions

**Fig. 1.** (Color online) Schematic presentation of hexagonal Ising nanowire. The blue and red spheres indicate magnetic atoms at the surface shell and core, respectively.

**Fig. 2.** (Color online) Time variations of the core and shell magnetizations ($m_C$ and $m_S$):

**(a)** Paramagnetic phase (p), $r=0.3$, $\Delta_S=-0.95$, $h_0$ = 0.78, and T = 0.6.

**(b)** Ferromagnetic phase (f), $r=4.0$, $\Delta_S=-0.5$, $h_0$ = 1.5, and T = 1.5.

**(c)** Antiferromagnetic phase (af), $r=-1.0$, $\Delta_S=0.0$, $h_0$ = 1.20, and T = 0.8.

**(d)** Nonmagnetic phase (nm), $r=0.2$, $\Delta_S=-5.0$, $h_0$ = 0.4, and T = 0.1.

**(e)** Mixed phase (nm+p), $r=0.1$, $\Delta_S=3.0$, $h_0$ = 6.0, and T = 0.1.

**Fig. 3.** (Color online) Thermal behaviors of the dynamic core and shell magnetizations with the various values of r and $\Delta_S$.

**(a)** r = 1.0, $\Delta_S = 0.0$, and $h_0$ = 0.1.

**(b)** r = - 0.5, $\Delta_S = -0.96$, and $h_0$ = 0.12.

**(c)** and **(d)** r = 0.1, $\Delta_S = 3.0$, and $h_0$ = 6.05.

**Fig. 4.** The dynamic phase diagrams in (h -T) plane of the hexagonal Ising nanowire. Dashed and solid lines show the first- and second-order phase transitions, respectively. The tricritical points are indicated with filled circles.

**(a)** r = 0.3 and $\Delta_S = -0.95$; **(b)** r = 4.0 and $\Delta_S = -0.5$; **(c)** r = 1.0 and $\Delta_S = 0.0$;

**(d)** r = -1.0 and $\Delta_S = 0.0$; **(e)** r = 0.1 and $\Delta_S = 3.0$; **(f)** r = 0.2 and $\Delta_S = -5.0$;

**(g)** r = -0.5 and $\Delta_S = -3.0$; **(h)** r = - 0.5 and $\Delta_S = -0.6$.

Figure 1

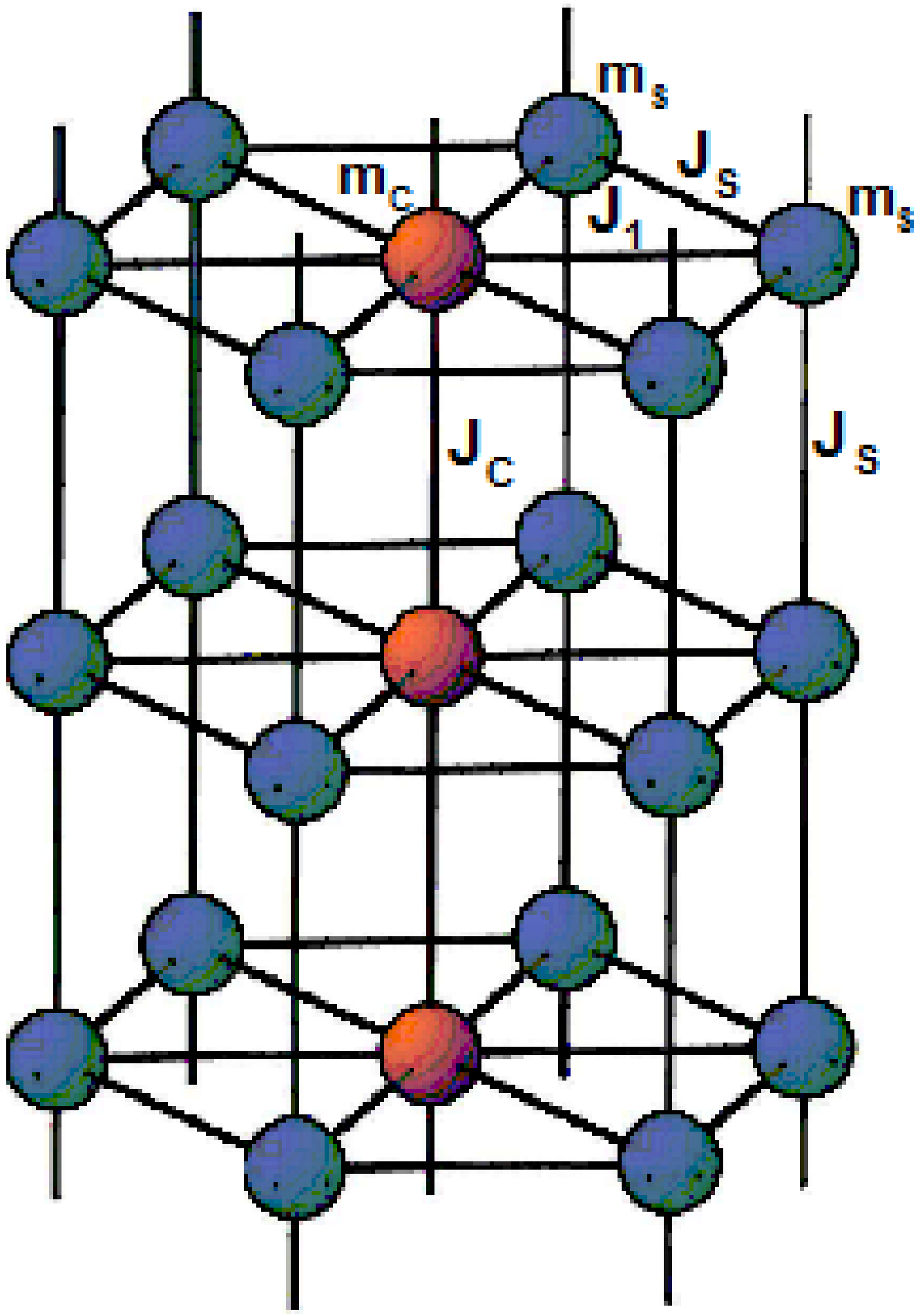

Fig. 1

Figure 2

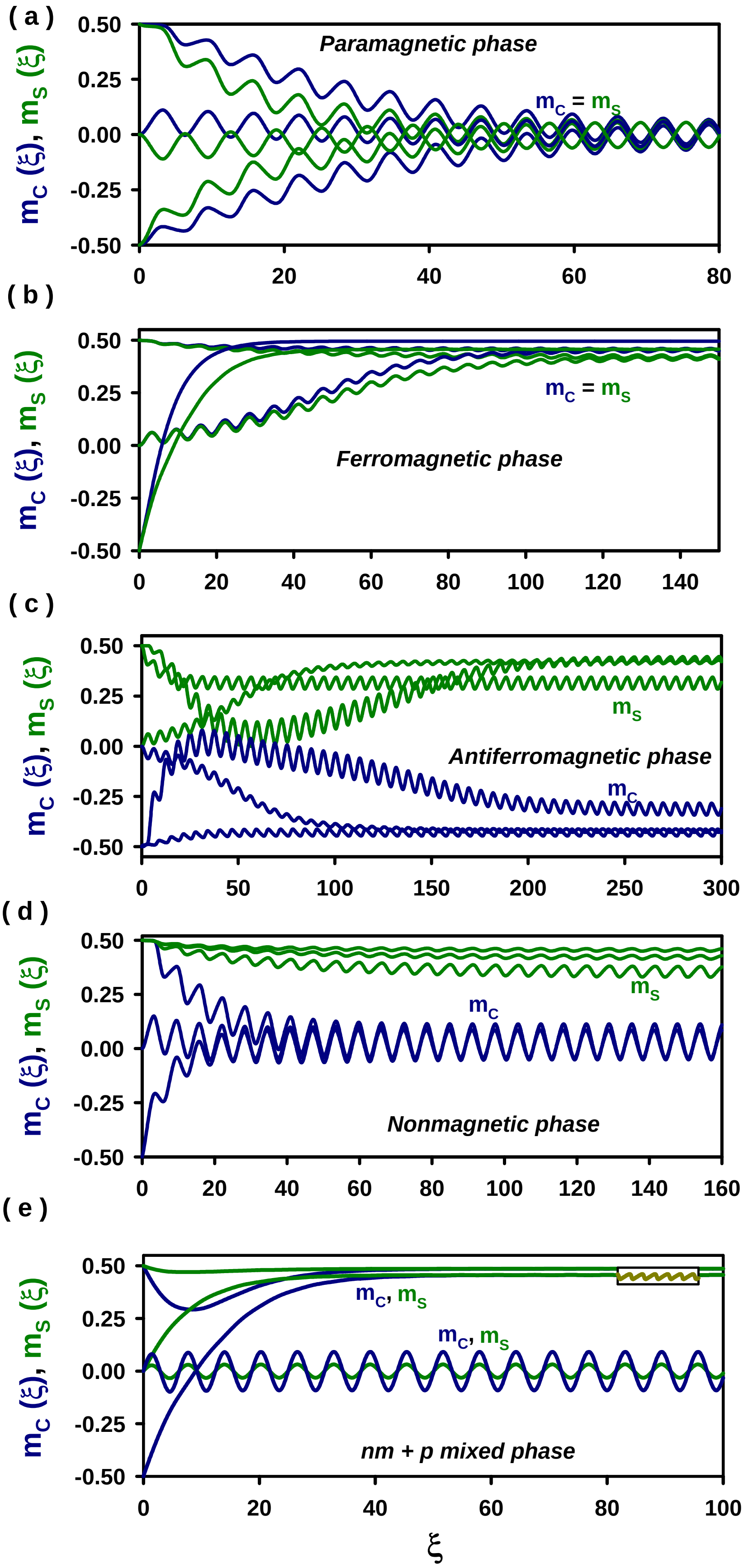

Fig. 2

Figure 3

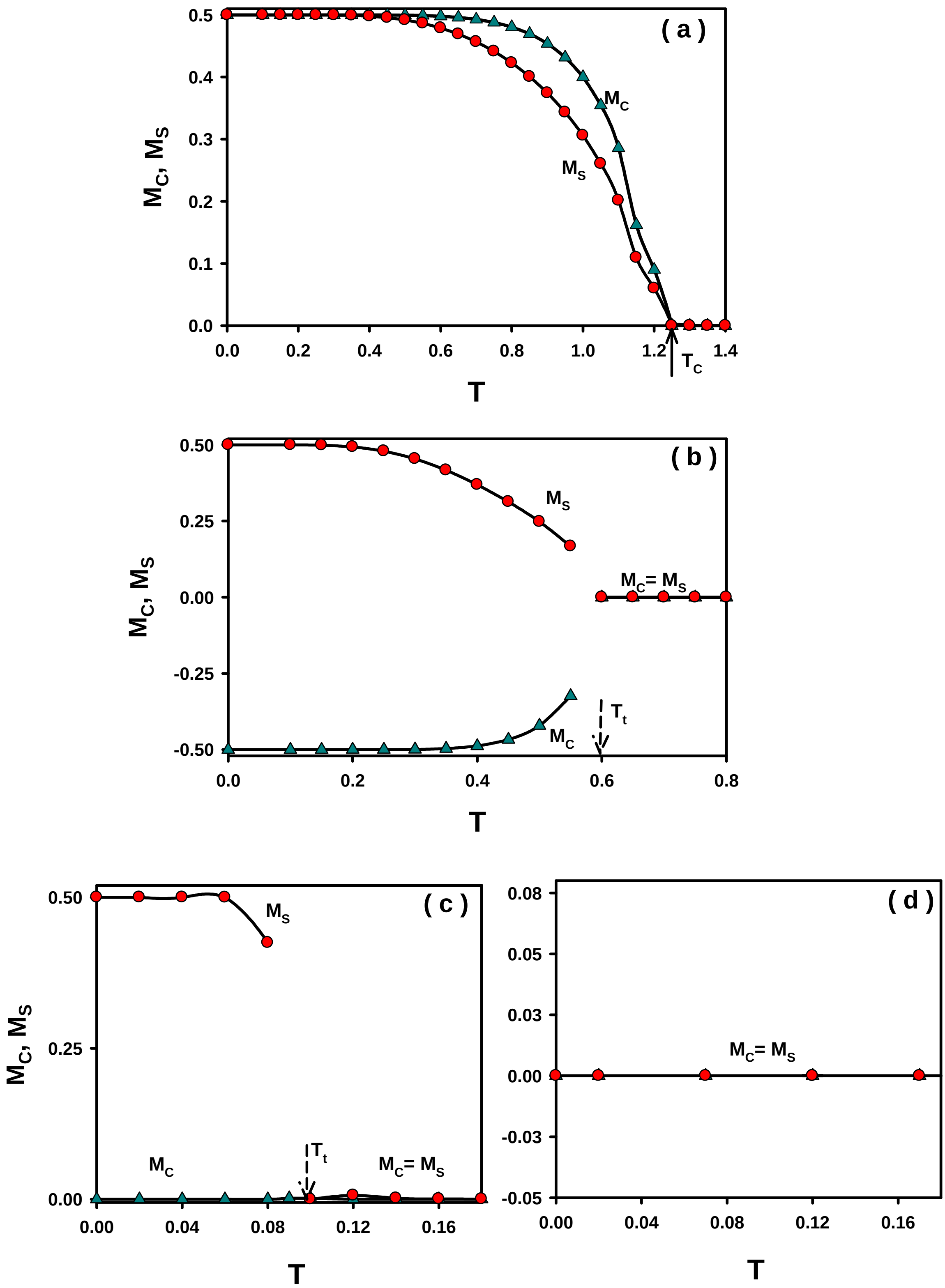

Fig. 3

Figure 4

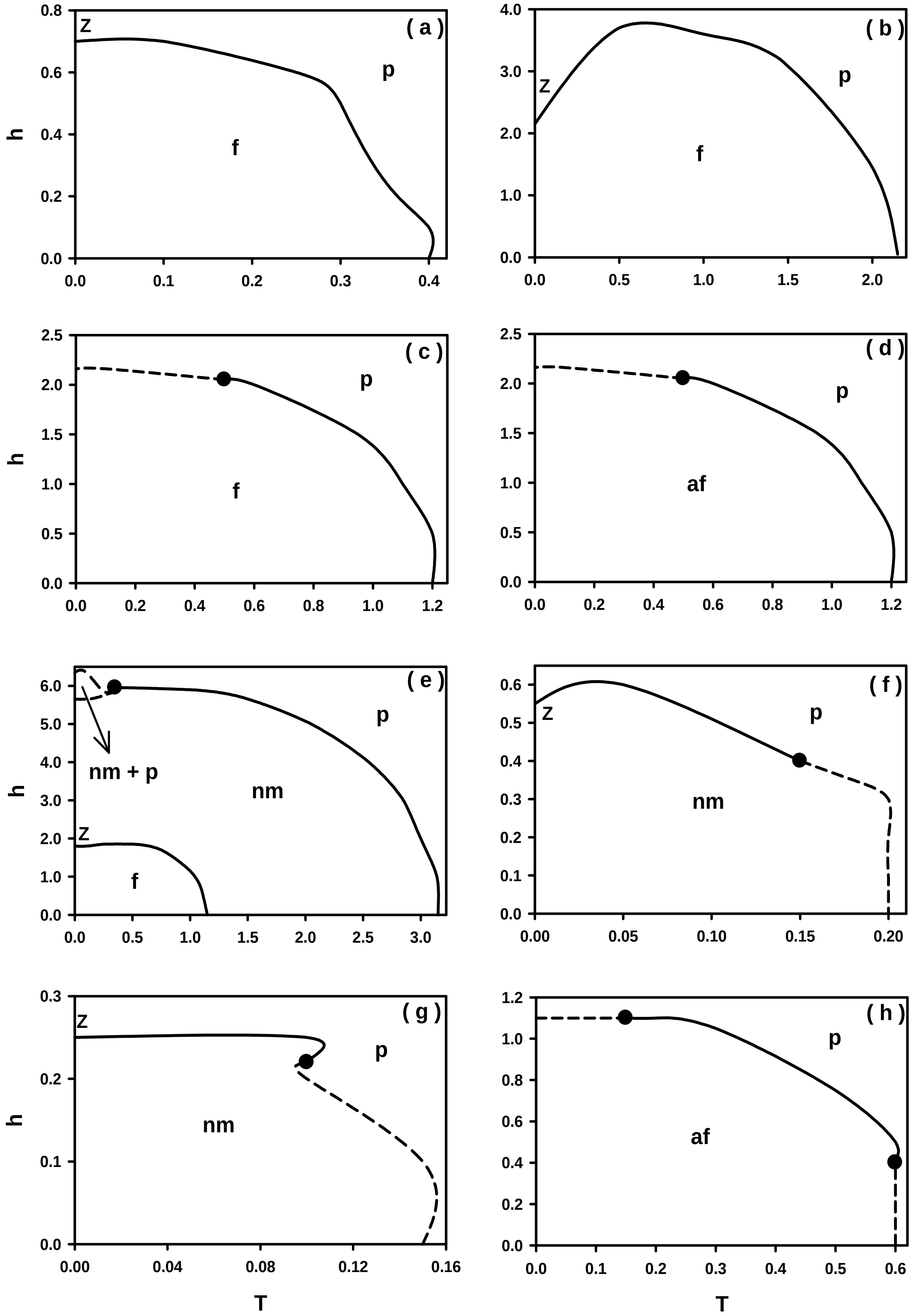

Fig. 4